\documentclass[preprint,aps,prc,showpacs,nofootinbib]{revtex4}

\usepackage{epsfig}
\usepackage{graphicx}

\newcommand\ba{\begin{eqnarray}}
\newcommand\ea{\end{eqnarray}}
\newcommand\nn{\nonumber}
\newcommand{\be}{\begin{equation}}
\newcommand{\ee}{\end{equation}}
\begin{document}

\title{Search for two photon exchange from $e^+ +e^-\to p+\bar p +\gamma$ data}

\author{E. Tomasi-Gustafsson}
\affiliation{\it DAPNIA/SPhN, CEA/Saclay, 91191 Gif-sur-Yvette Cedex,
France }
\author{E. A. Kuraev, S. Bakmaev}
\affiliation{\it JINR-BLTP, 141980 Dubna, Moscow region, Russian
Federation}

\author{S. Pacetti }

\affiliation{\it Centro Studi e Ricerche Enrico Fermi, Roma \\ and INFN, Laboratori Nazionali di Frascati, Italy}

\date{\today}

\begin{abstract}
We look for asymmetries in the angular distributions of events from recent data on $e^+ +e^-\to p +\bar p +\gamma$  from BABAR collaboration. From first principles, as the C-invariance of the electromagnetic interaction and the crossing symmetry, the presence of two-photon exchange would create a forward backward asymmetry in the data. The analysis of the available data shows no asymmetry, within an error of 2\%. This result is consistent with a structureless model for the proton, based on a calculation of 
$e^+ +e^-\to \mu^+ + \mu^- +\gamma$ with a proper replacement of the muon mass. As no systematic deviations are seen, we can conclude that these data do not give any hint of the presence of the two photon contribution, in all the considered kinematical range. 
\end{abstract}
\maketitle
The hadron structure is preferentially investigated using electromagnetic probes, assuming that the interaction occurs through one photon exchange.
The formalism of elastic, inelastic, polarized or unpolarized processes can be established in a transparent form, which involves explicitly the parametrization of the nucleon charge and magnetic distributions. However long ago, \cite{Gu73} it was pointed out that at large momentum transfer, the mechanism of two (or more) photon exchange can become important, in particular when the momentum is equally shared among the two photons. Recently this issue has been discussed as a possible explanation for the discrepancy among experimental data, in case of electron deuteron \cite{Re99} and electron proton elastic scattering \cite{twof}. As a model independent calculation is not feasible in case of hadronic targets, experimental evidence of such mechanism would be extremely important. If present, care should be taken in the analysis of different processes, and new, more complicated procedures and experiments would be unavoidable to extract information on the hadron structure. 

Model independent information on the presence of two photon exchange can be found in the angular distribution of the annihilation process
\be
e^+(p_+) +e^-(p_-)\to p(p_1)+\bar p(p_2).
\label{eq:eqr}
\ee 
In case of one photon exchange, the differential cross section for the case of unpolarized 
particles, is written as \cite{Du96}:  
\begin{equation}
\frac{d\sigma_{un}}{d\Omega } = \frac{\alpha^2\beta }{4t}D,~t=(p_+ + p_-)^2,
\label{eq:eq1}
\end{equation}
with
$$D=(1+\cos^2\theta )|G_M|^2+\frac{1}{\tau }
\sin^2\theta |G_E|^2, ~\tau=\frac{t}{4M^2 }, $$
where $M$ is the proton mass, $\theta $ is the angle between the momenta of the electron and the detected nucleon,  
in the $e^++e^-\rightarrow p+\bar p$ reaction center of mass frame (CMS), $\sqrt{t}=2E$ is the total CMS energy, $E$ is the incident energy. 
The even character of the distribution (\ref{eq:eq1}) is driven by the one-photon exchange mechanism. Two photon exchange would induce odd terms in $\cos\theta $. This is a model independent property of two photon exchange that has been proved in Ref. \cite{Re99}, and more recently in \cite{Re04}. 

Let us recall the arguments here. Assuming one-photon exchange, the conservation of the total angular momentum ${\cal J}$  allows one value: ${\cal J}=1$, and the quantum numbers of the photon: ${\cal J}^P=1^-$, $C=-1$. The selection rules with respect to the C and P-invariance allow two states for 
$e^+e^-$ (and $p\overline{p}$):
\begin{equation}
S=1,~\ell=0 \mbox{~and~} S=1,~\ell=2\mbox{~with~} {\cal J}^P=1^-,
\label{eq:tran}
\end{equation}
where $S$ is the total spin and $\ell$ is the orbital angular momentum. As a result the $\theta$-dependence of the cross section for $e^++e^-\to p+\overline{p}$, in the one-photon exchange mechanism is:
\begin{equation}
\displaystyle\frac{d\sigma}{d \Omega}(e^++e^-\to p+\overline{p})\simeq a(t)+b(t)\cos^2\theta, 
\label{eq:sig}
\end{equation}
where $a(t)$ and $b(t)$ are definite quadratic contributions of $G_{Ep}(t)$ and 
$G_{Mp}(t)$, $a(t),~b(t)\ge 0$ at $t\ge 4M^2$.

Such relation is equivalent in the space-like region, to the statement that the one-photon mechanism generates a linear $\cot^2{\theta_e/2}$ dependence of the Rosenbluth differential cross section for elastic $eN$-scattering 
($\theta_e $ is the electron scattering angle in Lab system).

Let us consider now the $\cos\theta$-dependence of the $1\gamma\bigotimes 2\gamma$-interference contribution to the differential cross section of  $e^++e^-\to p+\overline{p}$. The spin and parity of the $2\gamma$-states 
are not fixed, in general, but only a positive value of C-parity, $C(2\gamma)=+1$, is allowed.
An infinite number of  states with different quantum numbers can contribute, and their relative role is determined by the dynamics of the process $\gamma^*+\gamma^*\to  p+\overline{p}$, with both virtual photons.

But the $\cos\theta$-dependence of the contribution to the differential cross section for the $1\gamma\bigotimes 2\gamma$-interference can be predicted on the basis of its C-odd nature:
\begin{equation}
\displaystyle\frac{d\sigma^{(int)}}{d \Omega}(e^++e^-\to p+\overline{p})=\cos\theta[c_0(t)+c_1(t)\cos^2\theta+c_2(t)\cos^4\theta+...],
\label{eq:sig3}
\end{equation}
where $c_i(t)$, $i=0,1,..$ are real coefficients, which are functions of $t$,  only. This odd $\cos\theta$-dependence is essentially different from the even $\cos\theta$-dependence of the cross section for the one photon approximation, Eq. (\ref{eq:sig}).

Looking for C-odd contributions to experimental angular distributions for 
reaction (\ref{eq:eqr}), is therefore a model independent way to have a clear signal of the two photon mechanism.


Due to the luminosity, no data on angular distributions for 
reaction (\ref{eq:eqr}) are available with sufficient statistics. But recently it has been shown \cite{Ba06} that the emission of an initial photon in the process
\be
e^+ +e^-\to p+\bar p+\gamma 
\label{eq:eqrg}
\ee
can also give useful information on nucleon form factors. 

The angular distributions, have been published in Ref. \cite{Ba06}, for different ranges of the invariant mass of the $p\bar p$ pair,
$M_{p\bar p}$. Such distributions have been built with the help of a Monte Carlo (MC) simulation, which takes into account the properties of the detection and allows to subtract the background. 
The emission of an additional photon in the initial state induces a 'deformation' of the angular distribution of the final hadron, which is function of the angle of the photon and can be calculated in frame of QED. In the limit $Q^2\ll t$, the angular distribution for the reaction (\ref{eq:eqr}) can be parametrized as 
\be
\frac{dN}{d\cos\theta}=A\left [H_M(\cos\theta,M_{p\bar p} )+\left | \frac{G_E}{G_M}
\right |^2 H_E(\cos\theta,M_{p\bar p} )\right ]
\ee
where $A$ is an overall normalization factor, $H_{M,E}$ is a function, calculated with the help of MC.

Two different procedures have been applied in order to study the effect of radiative corrections: 
higher order radiative corrections have been calculated according to Ref. \cite{Ca97} and next-to-leading order radiative corrections in the initial state have been calculated according to Ref. \cite{Cz04}, and applied as a radiative correction factor to the mass spectrum. The consistency of the procedure has been checked at the level of percent.

Let us discuss the size and the origin of other possible radiative corrections, particularly of C-odd nature, which could contribute to an eventual asymmetry in the data.


Other odd contributions to the reaction (\ref{eq:eqr}), with respect to  $\cos\theta$, may arise due to $Z$-boson exchange and $C-odd$ interference of radiative amplitudes (including the emission of virtual and real photons). For moderate to large energies, but smaller than the $Z$-boson mass, $\sqrt{t}/M_Z\ll 1$, the $Z-$boson exchange can be neglected.
Its contribution can be evaluated of the order of ${\cal A}_Z\sim (t/M_Z)^2a_va_a\sim 10^{-6}$ where $a_v$ and $a_a$ are the vector and axial coupling constant of the $Z$ boson with the electron.

Let us define the asymmetry as:
\be
{\cal A}(c)=\displaystyle\frac{
\displaystyle\frac{d\sigma}{d\Omega}(c) -
\displaystyle\frac{d\sigma}{d\Omega}(-c)}
{\displaystyle\frac{d\sigma}{d\Omega}(c) +\displaystyle\frac{d\sigma}{d\Omega}(-c)},~c=\cos\theta. 
\label{eq:eqas}
\ee
Contributions arising from the initial state emission are canceled in leading logarithmic approximation (LLA) in the numerator, but they are enhanced in the denominator. We can write, in terms of structure functions:
\ba
&\displaystyle\frac{d\sigma}{d\Omega}(c) \pm \displaystyle\frac{d\sigma}{d\Omega}(-c)
\sim \int & dx_1{\cal D}(x_1,L){\cal D}(x_2,L)dx_2 
\left (1+\displaystyle\frac{\alpha}{\pi}K\right )\nn \\
&&
\left [d\sigma_B(p_-x_1,p_+x_2,c)\pm d\sigma_B(p_-x_1,p_+x_2,-c)\right ],
\label{eq:eq3}
\ea
where $x_1$ and $x_2$ are the energy fractions carried by the electron and the positron after emission of collinear photons.
In case of $e^+e^-\to \mu^+\mu^-$, this formula has been explicitly given in Ref. \cite{Ku91}. We can consider this calculation as a model for reaction (1), when a muon is a structureless proton.

One can write, for the Born cross section:
\be
\displaystyle\frac{
{d\sigma}_B}{d\Omega}(p_-,p_+)= \displaystyle\frac{\alpha^2}{4t}(1+c^2). 
\label{eq:eqb}
\ee
The emission of initial photon induces a shift in the kinematical invariants relevant to the process. The  shifted cross section can be written as:
\be
\displaystyle\frac{
{d\sigma}_B}{d\Omega}(x_1p_-,x_2p_+)= \displaystyle\frac{2\alpha^2}{t}
\displaystyle\frac{x_1^2(1-c)^2+x_2^2(1+c)^2}{[x_1+x_2-c(x_1-x_2)]^4}=
\displaystyle\frac{2\alpha^2}{t}F(x_1,x_2,c).
\label{eq:eqbs}
\ee
Due to the property $F(x_1,x_2,c)=F(x_2,x_1,-c)$, which holds in LLA, the numerator of (\ref{eq:eqas}) is identically zero, neglecting the non leading contributions which are included in the $K$ factor. 

\begin{figure}
\begin{center}
\includegraphics[width=18cm]{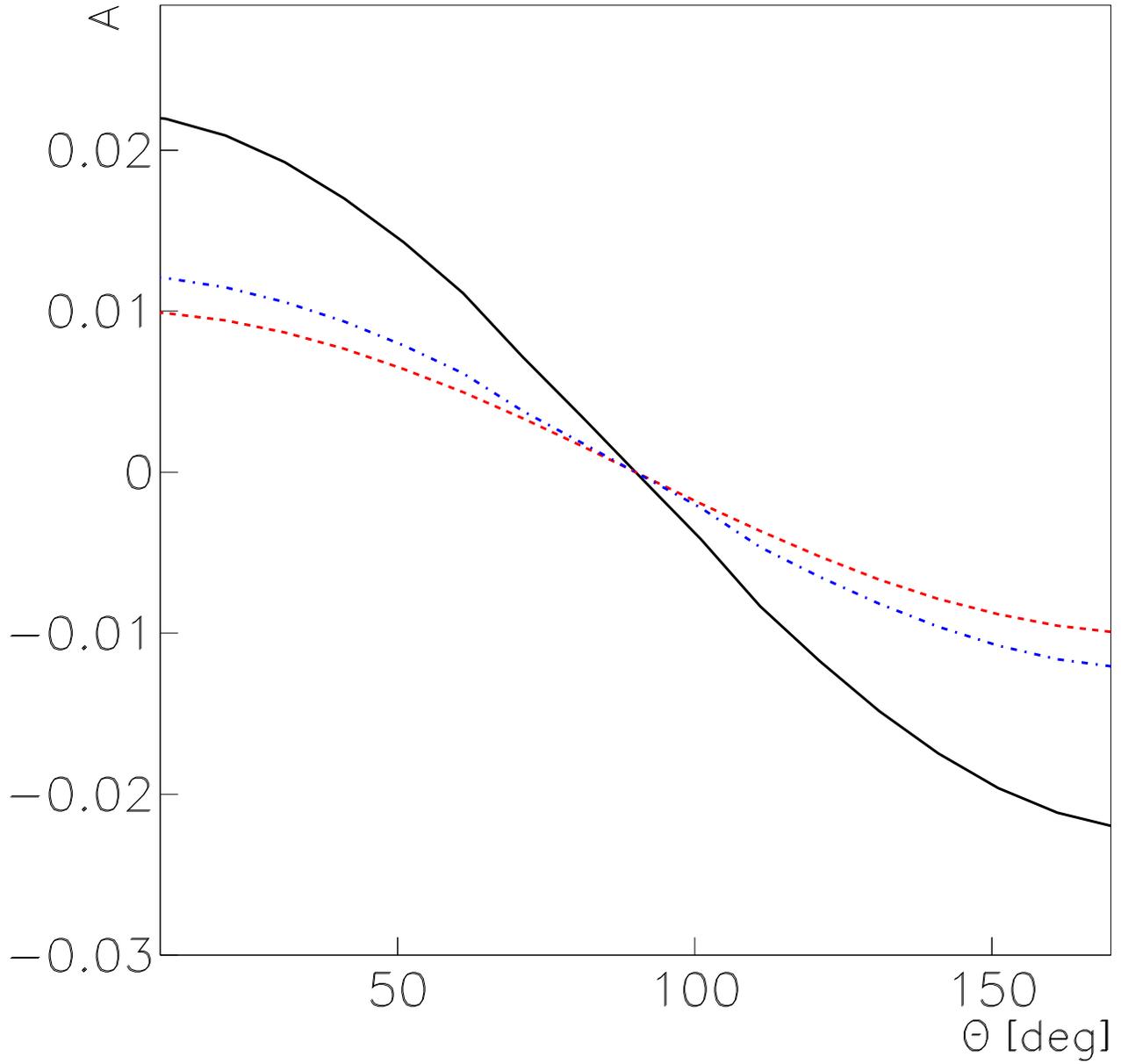}
\caption{\label{Fig:baba} Asymmetry due to soft photon emission
from \protect\cite{Ku91}, for s=4 GeV$^2$ : total contribution (solid line), energy dependent term calculated for $\Delta E/E=0.05$ (dashed line), energy dependent term subtracted (dash-dotted line).} 
\end{center}
\end{figure}

The non-leading contributions were calculated in Ref. \cite{Ku77}. Taking into account the emission of soft photons with energy $\omega\le\Delta E\ll\sqrt{t}/{2}$. For initial state radiation, one finds:
\be
\displaystyle\frac{d\sigma}{d\Omega}(c) + \displaystyle\frac{d\sigma}{d\Omega}(-c)=
2\displaystyle\frac{d\sigma_0}{d\Omega}\left [1+\displaystyle\frac{\alpha}{\pi}\left (\displaystyle\frac{3}{2}L
-2(L-1)\ln\displaystyle\frac{\Delta E}{E}+\displaystyle\frac{\pi^2}{3}-2\right )\right ],~L=\ln\displaystyle\frac{t}{m^2},
\label{eq:eqnl}
\ee
$m$ is the electron mass.
The polarization of vacuum should be included, but does not contribute to the  asymmetry, as it cancels in the ratio (\ref{eq:eqas}).
Typical value of such corrections is of the order of 2\% for quasi elastic annihilation (Fig. \ref{Fig:baba}). The largest contribution to the asymmetry is represented by a factor which depends on the soft photon energy $\Delta E$:
\be
A^{soft} (E)\simeq \frac {2\alpha}{\pi}\left (\ln \frac{1+\beta c}{1-\beta c} \ln \frac{E}{\Delta E}\right ),
\label{eq:eqsoft}
\ee
where $\beta$ is the CMS proton velocity, $\beta=\sqrt{1-4M^2/t}$. The C-odd soft contribution is shown in Fig. \ref{Fig:baba}, (solid line) as well as the energy dependent term   
(\ref{eq:eqsoft}) (dashed line) and their difference (dash-dotted line), for $t$=4 GeV$^2$ and $\Delta E/E=0.05$. The difference is especially instructive, because the energy dependent factor is compensated by hard photon emission, with energy $\omega>\Delta E$ . The hard contribution to the asymmetry $A^{hard}$ is explicitely calculated in Ref. \cite{Ku77}.  
Therefore we can expect a total contribution to the asymmetry  
$$A^{tot}=A^{soft}+A^{hard}=\frac {2\alpha}{\pi} \psi(c,\beta),~ |A^{tot}|\le 2\%,$$
where $\psi(c,\beta)$ is a complicated function of $c$ and $\beta$ with the following 
properties $|\psi(c,\beta)| \sim 1$ and with odd character with respect to $c$  \protect\cite{Ku91}. 

Radiative corrections have been calculated in Ref. \cite{Ku77} for electron and positron annihilation into two muons taking into account hard and soft photon emission, for asymptotic conditions, i.e. $\beta \approx 1$. Such kinematics represents an upper limit for the case under consideration. 

On the basis of this discussion, we can expect a total contribution from radiative corrections to the angular asymmetry not exceeding 2\%. In principle radiative corrections have been applied to the data, therefore part or all of the asymmetry arising by soft and hard photon emission should be already accounted for in the differential cross section. 

For each $\cos\theta$ bin, for each invariant mass interval, we calculate here the asymmetry defined as in Eq. (\ref{eq:eqas}). By definition, ${\cal A}(\cos\theta=0)$ vanishes.

\begin{figure}
\begin{center}
\includegraphics[width=18cm]{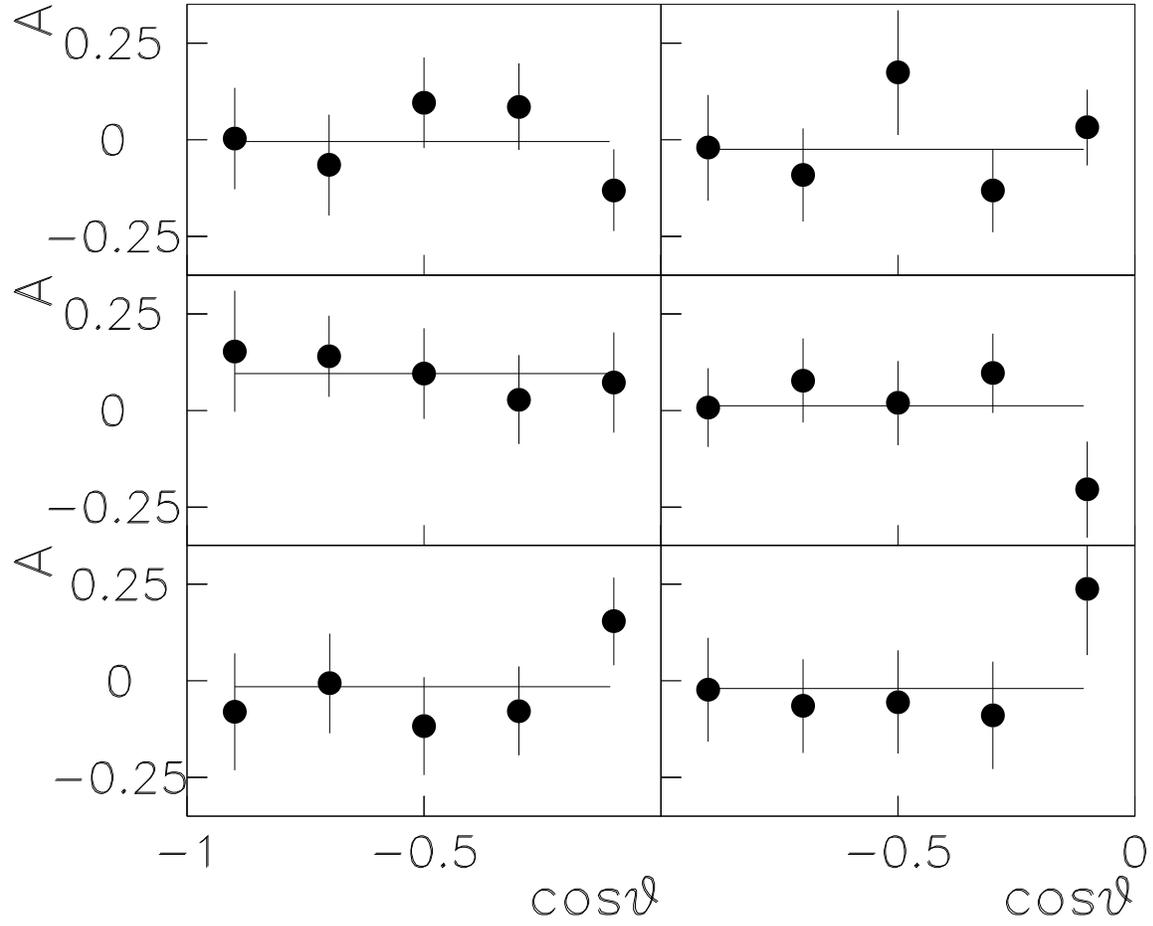}
\caption{\label{Fig:fig2}Forward backward asymmetry for 5 different 
$\cos\theta$ bins as a function of $\cos\theta$, and for different ranges of $t$.}
\end{center}
\end{figure}

\begin{figure}
\begin{center}
\includegraphics[width=10cm]{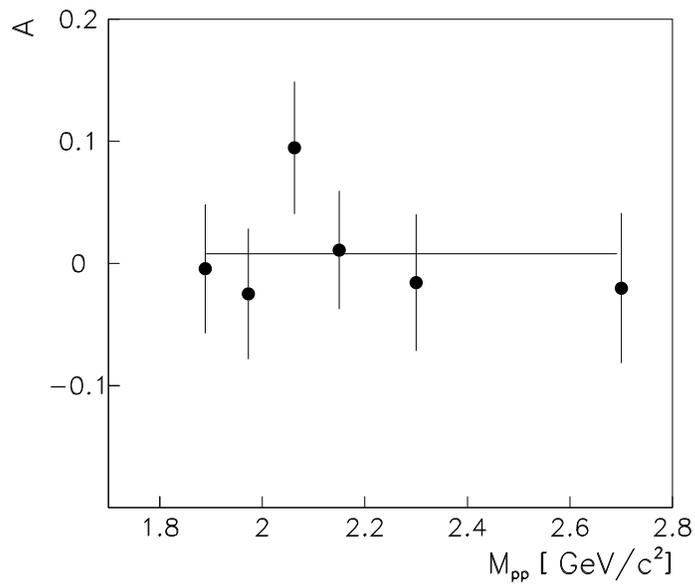}
\caption{\label{Fig:fig3}Average forward backward asymmetry as a function of $M_{p\bar p}$.}
\end{center}
\end{figure}

The dependence of the asymmetry as a function of $\cos\theta$ is rather flat and can be fitted by a constant. The experimental values as well as the results of the fit for the different bins of $M_{p\bar p}$ are shown in Fig. \ref{Fig:fig2} and in Table 1, together with the $\chi^2$ and the available number of points. The asymmetry is always compatible with zero and the typical error is $\sim 5\%$. As no systematic effect over $M_{p\bar p}$ appears, one can calculate the average for all the available range. The final average ${\cal A}=0.01 \pm 0.02$ is illustrated in Fig. \ref{Fig:fig3}. 

\begin{table*}
\begin{tabular}{|c|c|c|c|}
\hline\hline
$M_{p\bar p} $[GeV]& $A\pm \Delta A$ &
$\chi^2$& N \\
\hline\hline
1.877 - 1.900  & $-0.004\pm 0.053$ & 0.75 & 5\\
1.920 - 2.025  & $-0.025 \pm  0.053$ & 0.78 & 5\\
2.025 - 2.100  &  $0.095 \pm  0.054$ & 0.17& 5\\
2.100 - 2.200  & $ 0.011 \pm 0.048$ & 1.01& 5\\
2.200 - 2.400  & $-0.016 \pm  0.056$ & 0.84& 5\\
2.400 - 3.000  & $-0.020 \pm  0.061$ & 0.68& 5\\
\hline
 Average       &  $0.01\pm  0.02 $ &     0.67& 6\\
 \hline\hline
\end{tabular}

\caption{ Result of the fits for the forward backward asymmetry for different ranges of $M_{p\bar p}$.}
\label{tab1}
\end{table*}
One can conclude that the data do not show evidence for the presence of the two photon contribution at the level of their precision. This analysis is conceptually equivalent to the search of non linearities in the Rosenbluth fit which was done for elastic $ep$ scattering in Ref. \cite{ETG}, also with negative result. Two photon contribution is expected to become larger when the momentum transfer increases. Its study in the kinematical range covered by the present experiments requires more precise and dedicated measurements. 

\section{acknowledgments}
Thanks are due to Marco Zito and Rinaldo Baldini-Ferroli for their interest in this work and for fruitful discussions. 


{}

\end{document}